\PassOptionsToPackage{pdftex,final}{graphicx} 
\PassOptionsToPackage{pdftex,bookmarks=true}{hyperref}
\documentclass[final,submission,copyright,creativecommons]{eptcs}
\providecommand{\event}{ITRS 2014} 
\newif\ifwoes\woestrue
\newif\ifrecursion\recursionfalse
\newif\ifcolorpoint\colorpointtrue
\newif\ifbinary\binaryfalse
\usepackage[utf8x]{inputenc}
\usepackage[T1]{fontenc}
\usepackage{times}
\usepackage{proof}
\newtheorem{definition}{Definition}
\newtheorem{lemma}[definition]{Lemma}
\newtheorem{theorem}[definition]{Theorem}

\usepackage{lucretia}
\usepackage[pdftex,bookmarks=true]{hyperref}
\usepackage{graphicx}
\usepackage{wrapfig}
\usepackage{float}
\floatstyle{boxed}
\restylefloat{figure}
\usepackage[final]{listingsutf8}
\lstdefinelanguage{Lucretia}{
  keywords={let, new, true, false, func, return, if, do, then, else, ifhasattr},
  keywordstyle=\color{blue}\bfseries,
  identifierstyle=\color{black},
  sensitive=false,
  comment=[l]{//},
  morecomment=[s]{/*}{*/},
  commentstyle=\color{gray}\slshape\ttfamily,
  stringstyle=\color{red}\ttfamily,
  morestring=[b]',
  morestring=[b]"
}

\makeatletter
\def\@listI{\leftmargin\leftmargini \parsep 1pt plus 1pt minus 1pt%
\itemsep 1pt plus 2pt minus 1pt}
\let\@listi\@listI
\makeatother
\hypersetup{
  pdfauthor = {},
  pdfsubject = {F.3.3, F.4.1.},
  pdfkeywords = {types, scripting languages},
  pdfcreator = {LaTeX with hyperref package},
  pdfproducer = {pdflatex},
  bookmarksdepth=3,
  bookmarks=true,
  bookmarksopen=true,
  bookmarksopenlevel=3,
  final
}

\newcommand{\Lucretia}{Lucretia\xspace}

\title{\Lucretia\,---\, intersection type polymorphism for scripting languages%
  }
\author{
{Marcin Benke}
\institute{University of Warsaw%
 \thanks{This work was partly supported by the Polish government grant no N N206 355836.}
}
\email{ben@mimuw.edu.pl}
\and{Viviana Bono}
\institute{Dipartimento di Informatica dell'Università di Torino%
\thanks{This work was partly supported by the MIUR PRIN 2010-2011 CINA grant and by the ICT COST Action IC1201 BETTY.}
}
\email{bono@di.unito.it}
\and{Aleksy Schubert}
\institute{University of Warsaw%
$^*$
}\email{alx@mimuw.edu.pl}
}
\def\titlerunning{Lucretia}
\def\authorrunning{M. Benke, V. Bono, A. Schubert}

\begin{document}

\maketitle
\abovedisplayskip=6pt plus 4pt minus 6pt
\belowdisplayskip\abovedisplayskip

\let\delayfigures=\relax 
\def\comm#1{#1}
\def\syntaxfigure{
\begin{figure*}
\begin{displaymath}
\begin{array}{\comm{p{95pt}}rl@{~}l@{~}l}
  \comm{locations      &} \Loc\ni    & l \\
  \comm{variables      &} \Vars\ni   & x,y    & ::= & \mbox{ (identifiers) }\\
  \comm{value names    &} \VNames\ni & z,w      & ::= & x \mid l\\
  \comm{field names    &} \Fnames\ni & n,m    & ::= & \mbox{ (identifiers) }\\
  \comm{constants      &} \OConst\ni & c      & ::= & \mbox{ (literals)}\\
  \comm{function value &} \FVals\ni  & v_f    & ::= & \multicolumn{1}{p{120pt}}{
                                                       $\fdecl{x_1,\cdots,x_n}{t}{e}$}\\
  \comm{values         &} \Vals\ni   & v      & ::= & c \mid v_f \mid l\\
  \comm{function expressions &}       & e_f    & ::= & x \mid v_f \\
  \comm{atomic expressions &}        & a   & ::= & v \mid z \\ 
  \comm{expressions &} \Expr\ni   & e      & ::= & a \mid op_n(a_1,\cdots, a_n)\\
  \comm{               &}            &        &     &   \mid \knew
                                                        \mid a.n \mid a_1.n = a_2\\
   \comm{              &}            &        &     &   \mid \letin{x}{e_1}{e_2}\\
   \comm{              &}            &        &     &   \mid \ite{a}{e_1}{e_2}\\
   \comm{              &}            &        &     &   \mid e_f(a_1,\cdots, a_n)\\
   \comm{              &}            &        &     & \mid \iha{a}{n}{e_1}{e_2}\\
  \comm{objects        &} \Obj\ni    & o      & ::= & \{ \} \mid \{ L_f \}\\
  \comm{fields list    &}            & L_f    & ::= & n\!:\! v \mid n\!:\! v, L\\
   \comm{stores        &} \Heaps\ni  & \sigma & ::= & \cdot \mid (l, o) \sigma\\
\end{array}
\end{displaymath}
\caption{Abstract syntax}
\label{fig:syntax}
\end{figure*}
}

\def\sacsyntaxfigure{
\begin{figure}
\def\comm##1{}
\begin{displaymath}
\begin{array}{\comm{p{95pt}}rl@{~}l@{~}l}
  \comm{variables      &} \Vars\ni   & x,y,z,w  & ::= & \mbox{ (identifiers) }\\
  \comm{field names    &} \Fnames\ni & n,m    & ::= & \mbox{ (identifiers) }\\
  \comm{constants      &} \OConst\ni & c      & ::= & \mbox{ (literals)}\\
  \comm{function value &} \FVals\ni  & v_f    & ::= & \multicolumn{1}{p{120pt}}{
                                                       $\fdecl{x_1,\cdots,x_n}{t}{e}$}\\
  \comm{values         &} \Vals\ni   & v      & ::= & c \mid v_f \\
  \comm{function expressions &}       & e_f    & ::= & x \mid v_f \\
  \comm{atomic expressions &}        & a   & ::= & v \mid x \\ 
  \comm{expressions &} \Expr\ni   & e      & ::= & a \mid op_n(a_1,\cdots, a_n)\mid \knew\\
  \multicolumn{4}{@{~\quad\quad~~}l}{                     
                                                     \mid a.n \mid a_1.n = a_2
                                                     \mid \letin{x}{e_1}{e_2}
                                                     \mid e_f(a_1,\cdots, a_n)}\\
  \multicolumn{4}{@{~\quad\quad~~}l}{                     \mid \ite{a}{e_1}{e_2}
                                                     \mid \iha{a}{n}{e_1}{e_2}}\\
\end{array}
\end{displaymath}
\caption{Abstract syntax}
\label{fig:syntax}
\end{figure}
}

\def\semanticsfigure{
\begin{figure}[tb]
\small
\begin{displaymath}
\begin{array}{l@{~~}l}
  (\mbox{Let-Enter}) & \sigma, E\langle \letin{x}{e_1}{e_2}\rangle \leadsto \\
&\leadsto                         \sigma, E[\bullet:= \letin{x}{\langle e_1\rangle}%
                                                      {e_2}]\\
  (\mbox{Let})       & \sigma, E[\bullet :=\letin{x}{\langle v\rangle}
                                                      {e}] \leadsto
                         \sigma, E\langle e[x:=v]\rangle\\[1.4em]

  (\mbox{Op-Eval})      & \sigma, E\langle op_n(v_1,\cdots, v_n)\rangle \leadsto
                         \sigma, E\langle \delta_n(op_n, v_1,\cdots,v_n)\rangle\\[1.4em]

  (\beta v)            & \sigma, E\langle \fdecl{x_1,\cdots, x_n}{t}{e}(v_1,\cdots, v_n)\rangle \leadsto\\
                       & \leadsto \sigma, E\langle e[x_1:=v_1,\cdots,x_n:=v_n]\rangle\\[1.4em]

  (\mbox{If-True})   & \sigma, E\langle\ite{ \ktrue}{e_2}{e_3}\rangle \leadsto
                         \sigma, E\langle e_2\rangle\\
  (\mbox{If-False})   & \sigma, E\langle\ite{\kfalse}{e_2}{e_3}\rangle \leadsto
                         \sigma, E\langle e_3\rangle\\[1.4em]

  (\mbox{Ifhtr-True} )  & \sigma, E\langle\iha{l}{n}{e_1}{e_2}\rangle \leadsto
                          \sigma, E\langle e_1\rangle\\
                        & ~~~\mbox{ when } a\in\dom{\sigma(l)}\\
  (\mbox{Ifhtr-False} ) & \sigma, E\langle\iha{l}{n}{e_1}{e_2}\rangle \leadsto
                          \sigma, E\langle e_2\rangle \\
                        & ~~~\mbox{ when } a\not\in\dom{\sigma(l)}\\[1.4em]

  (\mbox{New})         & \sigma, E\langle\knew\rangle \leadsto
                         (l, \{\})\sigma, E\langle l\rangle \quad l \mbox{
                         fresh }\\
  (\mbox{SetAttr})      & \sigma, E\langle l.n=v\rangle \leadsto
                         \sigma [l:=\sigma(l)[n:=v]], E\langle v\rangle\\
                         
  (\mbox{GetAttr})       & \sigma, E\langle l.n\rangle \leadsto
                         \sigma, E\langle \sigma(l)(n)\rangle \quad\mbox{ when } n\in\dom{\sigma(l)}\\

\end{array}
\end{displaymath}
\caption{Semantic rules of \Lucretia}
\label{fig:semantics}
\end{figure}
}

\def\semanticsaltfigure{
\begin{figure}[tb]
\small
\begin{displaymath}
\begin{array}{l@{~~}l}
  (\mbox{Let-Propag}) & \mbox{if } \sigma, e_1 \leadsto \sigma', e_1' \\
                                   & \mbox{then } \sigma, \letin{x}{e_1}{e_2} \leadsto \sigma', \letin{x}{e_1'}{e_2}\\

  (\mbox{Let-Reduce})       & \sigma, \letin{x}{v}
                                                      {e} \leadsto \sigma, e[x:=v]\\

 (\mbox{Env}) & \mbox{if } \sigma, e \leadsto \sigma', e' \mbox{ then } \sigma, E\langle e \rangle
                           \leadsto \sigma', E\langle e' \rangle\\[1.4em]

  (\mbox{Op-Eval})      & \sigma, E\langle op_n(v_1,\cdots, v_n)\rangle \leadsto
                         \sigma, E\langle \delta_n(op_n, v_1,\cdots,v_n)\rangle\\[1.4em]

  (\beta v)            & \sigma, E\langle \fdecl{x_1,\cdots, x_n}{t}{e}(v_1,\cdots, v_n)\rangle \leadsto\\
                       & \leadsto \sigma, E\langle e[x_1:=v_1,\cdots,x_n:=v_n]\rangle\\[1.4em]

  (\mbox{If-True})   & \sigma, E\langle\ite{ \ktrue}{e_2}{e_3}\rangle \leadsto
                         \sigma, E\langle e_2\rangle\\
  (\mbox{If-False})   & \sigma, E\langle\ite{\kfalse}{e_2}{e_3}\rangle \leadsto
                         \sigma, E\langle e_3\rangle\\[1.4em]

  (\mbox{Ifhtr-True} )  & \sigma, E\langle\iha{l}{n}{e_1}{e_2}\rangle \leadsto
                          \sigma, E\langle e_1\rangle\\
                        & ~~~\mbox{ when } a\in\dom{\sigma(l)}\\
  (\mbox{Ifhtr-False} ) & \sigma, E\langle\iha{l}{n}{e_1}{e_2}\rangle \leadsto
                          \sigma, E\langle e_2\rangle \\
                        & ~~~\mbox{ when } a\not\in\dom{\sigma(l)}\\[1.4em]

  (\mbox{New})         & \sigma, E\langle\knew\rangle \leadsto
                         (l, \{\})\sigma, E\langle l\rangle \quad l \mbox{
                         fresh }\\
  (\mbox{SetAttr})      & \sigma, E\langle l.n=v\rangle \leadsto
                         \sigma [l:=\sigma(l)[n:=v]], E\langle v\rangle\\
                         
  (\mbox{GetAttr})       & \sigma, E\langle l.n\rangle \leadsto
                         \sigma, E\langle \sigma(l)(n)\rangle \quad\mbox{ when } n\in\dom{\sigma(l)}\\

\end{array}
\end{displaymath}
\caption{Semantic rules of \Lucretia (alternative)}
\label{fig:semantics_alt}
\end{figure}
}

\def\semanticsnewfigure{
\begin{figure}[tb]
\small
\begin{displaymath}
\begin{array}{l@{~~}l}
  (\mbox{Let-Propag}) & \mbox{if } \sigma, e_1 \leadsto \sigma', e_1' 
                                   \mbox{ then } \sigma, \letin{x}{e_1}{e_2} \leadsto \sigma', \letin{x}{e_1'}{e_2}\\[0.5ex]
  (\mbox{Let-Reduce})       & \sigma, \letin{x}{v}
                                                      {e} \leadsto \sigma, e[x:=v]\\[1ex]


  (\mbox{Op-Eval})      & \sigma,  op_n(v_1,\cdots, v_n) \leadsto
                         \sigma,  \delta_n(op_n, v_1,\cdots,v_n)\\[1ex]

  (\beta v)            & \sigma,  \fdecl{x_1,\cdots,
    x_n}{t}{e}(v_1,\cdots, v_n) \leadsto
\sigma,  e[x_1:=v_1,\cdots,x_n:=v_n]\\[2ex]

  (\mbox{If-True})   & \sigma, \ite{ \ktrue}{e_2}{e_3} \leadsto
                         \sigma,  e_2\\
  (\mbox{If-False})   & \sigma, \ite{\kfalse}{e_2}{e_3} \leadsto
                         \sigma,  e_3\\[2ex]

  (\mbox{Ifhtr-True} )  & \sigma, \iha{l}{n}{e_1}{e_2} \leadsto
                          \sigma,  e_1
                         \mbox{ when } a\in\dom{\sigma(l)}\\
  (\mbox{Ifhtr-False} ) & \sigma, \iha{l}{n}{e_1}{e_2} \leadsto
                          \sigma,  e_2 
                          \mbox{ when } a\not\in\dom{\sigma(l)}\\[2ex]

  (\mbox{New})         & \sigma, \knew \leadsto
                         (l, \{\})\sigma,  l \quad l \mbox{
                         fresh }\\
  (\mbox{SetAttr})      & \sigma,  l.n=v \leadsto
                         \sigma [l:=\sigma(l)[n:=v]],  v\\
                         
  (\mbox{GetAttr})       & \sigma,  l.n \leadsto
                         \sigma,  \sigma(l)(n) \quad\mbox{ when } n\in\dom{\sigma(l)}\\

\end{array}
\end{displaymath}
\caption{Semantic rules of \Lucretia}
\label{fig:semantics_new}
\end{figure}
}

\def\typesyntaxfigure{
\begin{figure}
  \begin{displaymath}
    \begin{array}{rlcl@{}}
      \TConst\ni  & t_b      &  &(\kint,\kbool,\kstr,\ldots)\\
      \Types_c\ni & t_c    &::=& t_f \mid t_f \land t_c\\ 
      \Types_f\ni & t_f    &::=&
                  \atype{\Psi}{\overline{t}}{t}{\Psi} \mid \forall X.t_f\\
      \Types'\ni  & t,u      &::=&  t_b \mid X \mid t_c \mid t\lor t \\
      \Types_a'\ni & q    &::=& t \mid \bot \mid t \lor \bot \\
      \Recs\ni   & r      &::=&  \{ \overline{n:q} \} \mid \{\}\\
      \Constr\ni & \Psi   &::=&  X\matches r, \Psi \mid \emptyset
    \end{array}
  \end{displaymath}
\caption{Types}
\label{fig:types}
\end{figure}
}

\def\sactypesyntaxfigure{
\begin{figure}
  \begin{displaymath}
    \begin{array}{rlcl@{}}
      \TConst\ni  & \multicolumn{3}{l}{t_b \quad (\kint,\kbool,\kstr,\ldots) \qquad
      \TVars\ni     X       }\\
      \Types_c\ni & t_c    &::=& t_f \mid t_f \land t_c\\ 
      \Types_f\ni & t_f    &::=&
                  \atype{\Psi}{\overline{t}}{t}{\Psi} \mid \forall X.t_f\\
      \Types\ni  & t,u      &::=&  t_b \mid X \mid t_c \mid t\lor t \\
      \Types_a\ni & q    &::=& t \mid \bot \mid t \lor \bot \\
      \Recs\ni   & r      &::=&  \{ \overline{n:q} \} \mid \{\}\\
      \Constr\ni & \Psi   &::=&  X\matches r, \Psi \mid \emptyset
    \end{array}
  \end{displaymath}
\caption{Types}
\label{fig:types}
\end{figure}
}

\def\modelsfigure{
\begin{figure}
%
%

\begin{displaymath}
  \infer[\mruloconst]%
    {
      \sigma;\Gamma\models c : t_c
    }%
    {
      c\in\OConst
    &
      \type(c) = t_c
    }
\end{displaymath}



\begin{displaymath}
  \infer[\mrulloc]%
    {
      \sigma;\Gamma,l:X \models l:X
    }%
    {
      l\in\Loc  & l\in\dom{\sigma}
    }
\end{displaymath}
\begin{displaymath}
  \infer[\mrulfunc]%
    {
      \sigma;\Gamma \models (v,t):t
    }%
    {
      \begin{array}{cc}
        \multicolumn{2}{c}{\emptyset;\Gamma\vd v:t;\emptyset}\\
        v\in\FVals & t\in\Types
      \end{array}
    }
\end{displaymath}
\begin{displaymath}
  \begin{array}{l@{\qquad}l}
    \infer[\mrullor]
      {
        \sigma;\Gamma\models v: t_1 \lor t_2
      }{
        \sigma;\Gamma\models v: t_1
      } 
    &
    \infer[\mrulror]
      {
        \sigma;\Gamma\models v: t_1 \lor t_2
      }{
        \sigma;\Gamma\models v: t_2
      } 
  \end{array}
\end{displaymath}
\caption{The relation $\models$.}
\label{fig:models}
\end{figure}
}
\def\updatefigure{
\begin{figure}\small
\textbf{Record update}
\[
  \begin{array}{c}
    r\updconstr \{\} = r
  \\
    \{a:u,r\}\updconstr \{a:u'\} = \{a:u',r\}
  \\
    \{\somefields\}\updconstr \{a:u'\} 
     = \{a:u',\somefields\}\quad\mbox{if $a\not\in\overline{n}$}
  \\
    r\updconstr \{a:u'',r'\} 
     = (r\updconstr\{a:u''\})\updconstr r'
  \end{array}
\]

\textbf{Constraint update}
\begin{displaymath}
  \begin{array}{l}
    \Psi \updconstr \emptyset = \Psi\\
  
    \Psi,X\matches r \updconstr X\matches r' =
    \Psi,X\matches (r\updconstr r')\\
    
    \Psi,X\matches r \updconstr\Psi',X\matches r' =
    (\Psi\updconstr\Psi'),X\matches (r\updconstr r')\\
    
    \Psi \updconstr\Psi',X\matches r' =
    (\Psi\updconstr\Psi'),X\matches r' \quad \mbox{ if } X\not\in\dom{\Psi}
  \end{array}
\end{displaymath} 
\caption{The update operation $\updconstr$.}
\label{fig:update-constraints}
\end{figure}
}


\def\defupdatefigure{
\begin{figure}\small
In the following
  ${}^*$ should be understood as any of ${}^+$ and ${}^-$:
\begin{gather*}
(\Psi,X\matches r)[X\gets^* \{a\}] = \Psi,X\matches (r[X\gets^* \{a\}]) \\
\Psi [X\gets^* \{a\}] = \Psi \ \mbox{when $X\not\in\dom{\Psi}$}\\ 
\{ a:t, \somefields \}[X\gets^+\{a\}] =
            \{ a:t^+,\somefields \} \\
\{ a:t,\somefields \}[X\gets^-\{a\}] =
     \{ a:\bot,\somefields \} \\
%
(t)^+ = t\  \mbox{for $t\in\Types$} \qquad
(t\lor\bot)^+ = t^+
\end{gather*}

\caption{Definiteness update $\Psi[X\gets^+ \{a\}]$ and $\Psi[X\gets^- \{a\}]$.}
\label{fig:definitness-update}
\end{figure}
}

\def\vdcfigure{
\begin{figure}[ht]
\small
\begin{displaymath}
\begin{array}{c@{~~}c@{}}

\infer[\vdcrefl]{q\vdc q}{} 

&

\infer[\vdcror]{q \vdc q \lor q'}{}  


\\[2ex]

\infer[\vdcrrefl]{\{\}\vdc \{\}}{}
&
\infer[\vdccrefl]{\emptyset\vdc \emptyset}{}

\\[2ex]
\infer[\vdctrans]{q_1 \vdc q_3}{q_1 \vdc q_2 \quad q_2 \vdc q_3}
&
\infer[\vdccevolve]
  {\Psi_1\vdc\Psi_2,X\matches r}
  {
   \begin{array}{c}
     \Psi_1\vdc\Psi_2\\
     X\not\in\FTV{\Psi_1}
   \end{array}
  }

\\[1ex]

\multicolumn{2}{c}{
\infer[\vdcstruct]
  {\{b:q,\overline{n:u}\}\vdc \{b:q',\overline{n:u'}\}}
  {q \vdc q' &
   \{\overline{n:u}\}\vdc \{\overline{n:u'}\}
  }
}
\\[2ex]
\multicolumn{2}{c}{
\infer[\vdccstruct]
  {\Psi_1,X\matches r_1\vdc\Psi_2,X\matches r_2}
  {
    \begin{array}{c}
      r_1\vdc r_2 \quad   \Psi_1\vdc\Psi_2 \\
      X\not\in\FTV{\Psi_1,\Psi_2}\\
    \end{array}
  }
}
\\

\end{array}
\end{displaymath}
\caption{Order over constraints}
\label{fig:constraints-order}
\end{figure}
}

\def\typingrulesfigure{\typingrulesfiguresize{\small}}

\newcommand{\typingrulesfiguresize}[1]{%
\begin{figure*}
#1
\begin{displaymath}
  \begin{array}{c}
    {
      \Psi;\Gamma,z:t   \vd   z
        : t;\Psi
    }
    {}
  \ \vaccessrule 
  \qquad

  \infer[\newrule]%
    {
       \Psi;\Gamma   \vd  \knew
         : X;X\matchess{},\Psi
     }
     {
        X\not\in \FTV{\Psi,\Gamma}
     }

\\[1ex]
      \Psi;\Gamma \vd   c
        : t_c;\Psi
  \ \construle    
  
\quad

  \infer[\botrule]%
    {
       \Psi_1;\Gamma   \vd  e
         : t;X\matches\{\overline{n:u},m:\bot\},\Psi_2
     }
     {
       \begin{array}{c}
       \Psi_1;\Gamma   \vd  e
         : t;X\matches\{\overline{n:u}\},\Psi_2
       \\
        X\not\in \FTV{\Psi_1,\Gamma} \qquad  m\not\in \overline{n}
      \end{array}
     }
\\[2ex]
  \infer[\plusrule]{
         \Psi;\Gamma\vd  +(e_1, e_2) : t; \Psi_2
        }%
        {
          \begin{array}{cc}
            \Psi;\Gamma   \vd  e_1 : t_1;\Psi_1&
            \Psi_1;\Gamma   \vd  e_2 : t_2;\Psi_2\\
            t_1, t_2\vdc\cbool\lor\cint\lor\creal &
            t = t_1\lor t_2
          \end{array}
        }
\\[1ex]
  \infer[\raccrule]%
    {
      \Psi;\Gamma  \vd   z.m : t;\Psi
    }
    {
      \begin{array}{c}
        {\Psi;\Gamma   \vd   z : X;\Psi}\quad
      \\
        \Psi\ni X\matches\{m:t,\somefields\}
      \end{array}
    }

\quad
  \infer[\urule]%
{\Psi; \Gamma\vd z_1.m = z_2 : t;\Psi\updconstr X\matchess{m:t}}
{\Psi;\Gamma\vd z_1:X;\Psi
&\Psi;\Gamma\vd z_2:t;\Psi}

\\[1ex]

  \infer[\letrule]%
    {
      \Psi_1;\Gamma   \vd    \letin{x}{e_1}{e_0}
         : t;\Psi_3
    }
    {
      \begin{array}{c}
      \Psi_1;\Gamma   \vd    e_1
         : t_1;\Psi_2
      \\
      \Psi_2;\Gamma,x:t_1   \vd   e_0
         : t;\Psi_3
      \end{array}
    }

\quad
  \infer[\condrule]%
  {
    \Psi;\Gamma   \vd   \ite{a}{e_1}{e_2}
      : t;  \Psi' 
  }%
  {
    \begin{array}{c}
      {\Psi;\Gamma   \vd   a
        : \cbool;\Psi}\\
      \Psi;\Gamma \vd   e_i
       : t;\Psi'  \quad \mbox{ for } i=1,2 
    \end{array}
  }

\\[1ex]

  \infer[\vdcrule]%
  {
    \Psi;\Gamma\vd e:t;\Psi'_1
  }{%
    \Psi;\Gamma\vd e:t;\Psi_1 & \vdash\Psi_1\vdc\Psi'_1 
  }

  \quad

 \infer[\typevdcrule]%
    {
       \Psi_1;\Gamma   \vd  e: t';\Psi_2
     }
    {
       \Psi_1;\Gamma   \vd  e: t;\Psi_2 & t\vdc t'
    }
\\[2ex]
  \infer[\fdeclrule]%
    {
      \Psi;\Gamma  \vd  \fdecl{\overline{x}}{}{e} : t_f;\Psi
    }
    {
      \begin{array}{c}
      \Psi_s;\Gamma,\overline{x:s} \vd e:t;\Psi_t
    \\
     \overline{X} = \FTV{\overline{s},\Psi_s,t,\Psi_t}\setminus\FTV{\Gamma}
    \\
     t_f \equiv \forall\overline{X}\atype{\Psi_s}{\overline{s}}{t}{\Psi_t}
      \end{array}
    }
\\[2ex]
\infer[\fapprule]{
      \Psi;\Gamma   \vd   e_f(\overline{w})
        : \theta(r); \Psi\updconstr\theta_1(\Psi_r)  }%
{
      \begin{array}{c}
       \Psi;\Gamma    \vd   e_f
          : t_f;\Psi 
        \quad t_f \equiv \forall\overline{X}.
                 \atype{\Psi_s}%
                       {\overline{s}}{r}{\Psi_r}
         \\
        \theta:\overline{X}\to\overline{Y} \mbox{ is renaming}
         \qquad
         \FTV{t_f}\parallel \theta\\
        \FTV{\Psi}\cap(\dom{\theta(\Psi_r)}-\dom{\theta(\Psi_s)})=\emptyset\\
        \Psi;\Gamma  \vd   \overline{w}
          : \theta(\overline{s}); \Psi
         \\
         \Psi\vdc \Psi\updconstr\theta(\Psi_s)
         \qquad
         \dom{\theta(\Psi_s)}\subseteq\dom{\Psi}
      \end{array}
}
\\[1ex]
  \infer[\fcdecomp]%
    {
      \Psi;\Gamma  \vd  \fdecl{\overline{x}}{}{e} : t_1\land t_2;\Psi
    }
    {
      \begin{array}{c}
        \Psi;\Gamma  \vd  \fdecl{\overline{x}}{}{e} : t_1;\Psi\\
        \Psi;\Gamma  \vd  \fdecl{\overline{x}}{}{e} : t_2;\Psi
      \end{array}
    }
    \hskip 3pt plus 9pt minus2pt
    
  \infer[\ifhatrule]%
    {
      \Psi;\Gamma   \vd   \iha{a}{n}{e_1}{e_2}
         : t;\Psi_2
    }
    {
\begin{array}{c}
      \Psi;\Gamma   \vd   a:X;\Psi
    \\
      \Psi[X\gets^+ \{n\}];\Gamma   \vd   e_1
         : t;\Psi_2
    \\
      \Psi[X\gets^- \{n\}];\Gamma   \vd   e_2
         : t;\Psi_2
\end{array}
    }
\\[1ex]
   \infer[\parbox{24pt}{$\fdcomp{\mbox{i}}$}]%
    {
      \Psi_1;\Gamma   \vd   z
        : t_i; \Psi_2   }%
    {
      \begin{array}{c}
      \Psi_1;\Gamma   \vd   z
        : t_1\land t_2; \Psi_2
      \end{array}
    }
 \quad
  \infer[\ifhatrulete]
    {
        \Psi;\Gamma \vd   \iha{a}{n}{e_+}{e_-}
        : t;\Psi_2
    }%
    {
      \begin{array}{c}
      \Psi;\Gamma  \vd   a:X;\Psi
      \quad
      \Psi;\Gamma  \vd   e_*
         : t;\Psi_2
      \\
      \Psi[X\gets^* \{n\}]=\Psi \qquad
      *\in \{+, -\}
      \end{array}
    }
\relax
\end{array}
\end{displaymath}

\caption{Typing rules}
\label{fig:rules}
\end{figure*}
}

\ifx\delayfigures\undefined%
\syntaxfigure
\semanticsfigure
\typesyntaxfigure
\updatefigure
\typingrulesfigure
\vdcfigure
\defupdatefigure
\modelsfigure
\else\relax%
\fi

\begin{abstract}

  Scripting code may present maintenance problems in the
  long run. There is, then, the call for methodologies that make it
  possible to control the properties of programs written in dynamic languages in an automatic
  fashion. We introduce \Lucretia, a core language with an introspection
  primitive. \Lucretia is equipped with a (retrofitted) static type
  system based on local updates of types that describe the structure
  of objects being used. In this way, we deal with one of the most
  dynamic features of scripting languages, that is, the runtime
  modification of object interfaces. Judgements in our systems have a
  Hoare-like shape, as they have a precondition and a
  postcondition part. Preconditions describe static approximations of the interfaces of
  visible objects before a certain expression has been executed and
  postconditions describe them after its execution.  The field
  update operation complicates the issue of aliasing in the system.
  We cope with it by introducing intersection types in method
  signatures.

 \end{abstract}

\section{Introduction}
\label{sec-intro}
\hyphenation{meth-od-ol-o-gies} 
%
Dynamic languages optimise the programmer time, rather than the machine time, and
are very effective when small programs are constructed
\cite{Prechelt00,WrigstadEFNV09}. The advantages of the languages that help in development of short programs can be detrimental in the long run.
Succinct code, which has clear advantages over short-term programming,
gives less information on what a
particular portion of code is doing (and figuring this out is critical
for software maintenance, see
\cite{Sasso96,KoMCA06}). As a result, productivity of software
development can be in certain situations impaired \cite{MayerHRTS12}.
In particular, strong invariants a programmer can rely
on in understanding of statically typed code are no longer valid, e.g., the type of a
particular variable can easily change in an uncontrolled
way with each function call in the program.

Still, systems that handle complex and critical tasks such as the Swedish
pension system \cite{Stephenson01}, developed in Perl, are deployed and maintained.
Thus it is desirable to study methodologies which help programmers in
understanding their code and keeping it consistent. To this end,
\emph{retrofitted} type systems%
\footnote{A retrofitted type system is a a type system that was designed after the language. In particular, this is used in the setting of dynamic languages to indicate a static type system flexible enough to accept their most common idioms, that would be ill-typed with a classical type system, but that are run-time correct.
}
 may be an approach to bridge the gap between flexibility and type safety. 

Our proposal is
a retrofitted type system for a 
calculus with a reflection primitive. Our type system handles one of the most dynamic features of object-oriented scripting languages, the runtime modification of object interfaces. In particular, the runtime type of an
object variable may change in the course of program execution. This feature can be tackled to some extent through the introduction of a single assignment form for
local variables. Still, this cannot be applied easily to object fields. 
%
%
On the other hand, the information that statically describes the evolution of the runtime
type of a variable cannot be just a type in the traditional sense, but must reflect the journey of
the runtime type throughout the control flow graph of the program. However, it would
be very inconvenient to repeat the structure of the whole control flow graph for
each variable in the program. It makes more sense to describe the
type of each variable at program points which are statically available and this is the approach we follow in this paper. 
In our calculus, a variable referring to an object is annotated with a type variable paired
with a constraint expressing an approximation (a lower bound) of the actual type
of the object. Our type system design draws inspiration from the work on
type-and-effect systems \cite{Marino09,Gifford86,DBLP:books/daglib/0098135}.
We present our typings in a different manner, i.e.,
one where an effect is described by two sets of constraints that express
type approximations before and after execution of an instruction. The sets
of constraints together with the typed expression can be viewed as a triple in
a Hoare-style program logic.

An important element of the language design is the way functions (called methods
in object-oriented vocabulary) are handled. The function types describe 
contracts associated with the functions. 
We obtained a satisfactory level
of flexibility of function application due to type polymorphism. We use
two kinds of polymorphism here that serve two different purposes. The first
one is the parametric polymorphism, similar to the one of
System F. Through universal quantifier instantiation we make it possible to
adapt the function type to different sets of parameters. 
The second one is a form of ad-hoc polymorphism obtained through
the use of intersection types \cite{BarendregtCD83} and its purpose is to
provide particular contracts that are for specific aliasing schemes, i.e., one
may describe additional possible behaviours of a function that cannot be
described by instantiation of a universal type.

\newif\ifevolution
\evolutionfalse




\syntaxfigure

\section{Overview of the Calculus}
\label{sec:calculus}

The syntax of our calculus is depicted in Figure~\ref{fig:syntax}.
The elements of the set $\VNames=\Vars\cup\Loc$ are called {\em value
names}.
The calculus is object-based and our objects are records of
pairs {\em fieldname:value}. Moreover, it is imperative, that is, it has side-effects, therefore we have a heap where objects are stored.
Methods are modelled by fields containing functions.
There is no built-in concept of \emph{self}, but it can be encoded (see the examples in Section~\ref{sec:examples}). 
Values are either constants, functions,
locations (the latter do not appear in source programs, only in the
semantics).  


Expressions include 
value names,
primitive operation application, an object creation
operation, field access, field update, let-assignment, function application, a conditional
expression,  an
introspection-based conditional expression checking if a certain field
belongs to an object.

The operational semantics is presented in Figure~\ref{fig:semantics_new}.
The construct $\klet$ is the only possible evaluation context of the calculus, and rule (Let-Propag) takes care of the propagation of the reduction, while (Let-reduce) performs the appropriate substitution of the computed value $v$, once this is obtained. Rule (Op-Eval) applies the semantical
counterpart of the operation symbol to the given arguments. Rule
$(\beta_v)$ is the call-by-value function application. Rules (If-True)
and (If-False) are self-documented. Rules (Ifhtr-True) and
(Ifhtr-False) check whether a certain field belongs or not to an
object allocated in the heap, and choose a computation branch
accordingly. Rule (New) allocates a fresh address in the heap. Rule
(SetAttr): either adds the field $n$ to the object allocated at
location $l$, initialised with value $v$, if $n$ does not exist in
the object; or updates $n$ with $v$, otherwise. Rule (GetAttr)
extracts the value of the field $n$ from the object at location $l$, if
$n$ belongs to the object. Note that the semantics is
deterministic.

\semanticsnewfigure



The usage of an object field depends on its type, and since the type clearly depends on the computation flow, we need to update the constraints via static analysis of the computation flow; to keep track of the knowledge about the current fieldset, we use judgements which are a combination of usual typing judgements, and Hoare-style triples: $\Psi_1; \Gamma \vd e : t; \Psi_2$,
where $\Psi_i$ are constraint sets representing type information about the objects in expression $e$, respectively before and after considering the effects of expression. We call them the \emph{precondition} and the \emph{postcondition}. The type information associated with an expression is, then, a combination of two items: a representation of its actual type and a set of constraints on objects in the relevant part of the heap.

New fields can be added dynamically to our objects, moreover any existing field
can be assigned with values of different types during the computation, as it
happens in dynamic languages (e.g., Python, JavaScript, Ruby). An object type,
then, is not fixed once and forever. We decided, therefore, to type an object
with a \emph{constrained} type variable, written
$X\matches\{\somefields\}$, describing some type information for the listed
fields of an object of type $X$ 
(we write $\overline{\alpha}$ for a sequence $\alpha_1,\ldots, \alpha_{k}$).

One group of challenges in the design of the type system is posed by forks and
joins in the control flow.
Consider, for instance, \[
\ite{b}{x.n=1}{x.n=\mathtt{"hello"}} 
\]
Statically, we do not know whether $x$ has field $n$ of type $\kint$ or
of type $\kstring$. To keep track of both possibilities, we introduce
\emph{union} types: we type $x$ with type $X$, where $X\matches
\{n:\kint\lor\kstring\}$.
Another example is $\ite{b}{x.n=1}{0}$: statically, we do not know whether $x$
has field $n$, but if it does, it is of type $\kint$. To be able to track the
possible absence of a field, we introduce a \emph{bottom} type: we type $x$ with
type $X$, where $X\matchess{n:\kint\lor\bot}$. Moreover, the constraint
$X\matchess{n:\bot}$ means that the field is definitely absent.

Field access is allowed only if the types indicate the field is definitely present; we can then check whether $x$ has field $n$ as in 
\[\iha{x}{n}{x.n+1}{0}\] to decide whether it is possible to access $n$ or not.
%

We use intersection types to capture
possible different aliasing scenarios (cf. Section ~\ref{sec:examples}). 


\subsection{Types}
\label{sec:types}
\label{sec:types-types}

\sactypesyntaxfigure

The syntax of types is shown in Figure~\ref{fig:types}. 
We use an abbreviation $\{m:u,r\}$ for $\{m:u,\somefields\}$
where $r=\{\somefields\}$, $m\not\in\overline{n}$.
We impose additional, natural
restrictions on the shape of the records and constraints.
We require that in a record of the form $\{\somefields\}$ the labels in
$\overline{n}$ are unique.  
For a constraint $\Psi = \overline{X\matches r}$ we require that $r\in\Recs$
and that the variables $\overline{X}$ are also unique.

The shape of all types but function types is self-explanatory. A function type is made of: domain information, that is, the type of its arguments and a set of constraints that can be read as preconditions to the function application; and codomain information, the return type and a set of constraints which are the postconditions holding after the function body has been executed.
  

We say that $m\in\dom{\{\somefields\}}$ when $m$ is an element of
$\overline{n}$. Similarly, we say that $Y\in\dom{\Psi}$ when
$\Psi = \overline{X\matches r}$ and $Y$ is one of the elements of
$\overline{X}$. We define the set of free variables $\FTV{\Psi}$ in a set of
constraints $\Psi$ so that when
$\Psi = X\matches r, \Psi'$ we have $X\in\FTV{\Psi}$,
$\FTV{r}\subseteq\FTV{\Psi}$ and $\FTV{\Psi'}\subseteq\FTV{\Psi}$. Moreover,
we consider $\forall$ to be a binding operator so that
$\FTV{\forall X.t} = \FTV{t}-\{X\}$.

Judgements are of the form $\Psi_1;\Gamma \vd   e: t;\Psi_2$, where $e$ is an expression, $t$ is a type,
$\Gamma$ is an environment, and $\Psi_1$ and $\Psi_2$ are type variable constraint sets, as described earlier.

We use type variable renaming, indicated with $\theta$, to adapt universally quantified types
to different situations they can be used in. 

Its formal definition
follows. 

\begin{definition}[Renaming]
  \label{df:renaming}
  A bijection $\theta:{\cal X}\to{\cal Y}$ where ${\cal X}\cup{\cal Y}$ is
  a finite subset of $\Vars$ is called \emph{renaming}.
  We extend it structurally to types, expressions,
  environments and constraints with avoiding name clashes for bound
  variables. We use the notation $\dom{\theta} = {\cal X}$ and $\img{\theta}={\cal Y}$. 
  When sequences $\overline{X}, \overline{Y}$ of unique variables
  have the same length we write $[\overline{X}:=\overline{Y}]$ for a
  renaming $\theta$ such that $\theta(X_i)=Y_i$ for $X_i\in\overline{X}$. We assume
  that
  $\theta(Z)=Z$ for $Z\not\in\overline{X}$. We apply
  $[\overline{X}:=\overline{Y}]$ as a suffix, i.e.
  $t[\overline{X}:=\overline{Y}] = \theta(t)$.
  We write $A\parallel \theta$ when
  $A\cap(\img{\theta}-\dom{\theta}) = \emptyset$.
\end{definition}

\noindent
Observe that these renamings, unlike type instantiation in System F, 
cannot substitute two universally quantified variables with the same variable.
This is an important design choice as we believe that the form
of types should not hide other information.
The standard convention that makes it possible to glue together two different variables
puts on type readers the burden of
checking if different uniting schemes do not lead to unexpected situations, that is, unexpected aliasing, in our case.

\ifwoes
\subsection{Weakening Woes}
\label{sec:weakening}

Since the type information changes with the control flow,
a constraint update operation plays a central role in our system.
For compositionality, the following ``knowledge monotonicity'' with
respect to the constraint update operation must hold. 



\myparagraph{Monotonicity principle}
{\it
For every set of constraints $\Psi$ and derivable judgement $\Psi_1;\Gamma\vd e:t; \Psi_2$,
 such that variable names for objects created in $e$ are fresh with respect to $\Psi$, we can derive
 \[\Psi\updconstr\Psi_1;\Gamma\vd e:t; \Psi\updconstr\Psi_2\] 
}

Intuitively $\Psi\updconstr\Psi_1$ means the set of constraints $\Psi$
is updated with constraints from $\Psi_1$; it is formally defined in
Figure~\ref{fig:update-constraints}.

We observe that our conditional typing rules must have the same postconditions for the two branches (see rules 
$\condrule$ and rule $\ifhatrule$ in Figure~\ref{fig:rules}).
In Hoare logic, equalising branches' postconditions is obtained via
weakening, which in our case might be formulated more or less like
this:
\[
\infer{\Psi_1;\Gamma\vd e:t; \Psi}
{\Psi_1;\Gamma\vd e:t; \Psi' & \Psi'\vdc \Psi}
\]
where $\Psi'\vdc\Psi$ means that $\Psi$ is weaker than $\Psi'$.
We need, however, to be careful that weakening obeys monotonicity, lest the system be unsound (we have the scars to show for it).
 
One example of weakening pitfall is forgetting a constraint, i.e.:
\[\Psi,X\matches r \vdc\Psi\]
Let's say we can infer
\[X\matchess{};\Gamma\vd x.m = 1 : \kint ;X\matchess{m:\kint}\]
Forgetting the constraint would allow us to infer 
\[ X\matchess{};\Gamma\vd x.m = 1 : \kint ;\emptyset \]
while monotonicity with $\Psi=X\matchess{m:\kstr}$ requires that
\[ X\matchess{m:\kstr};\Gamma\vd x.m = 1 : \kint ;X\matchess{m:\kstr} \]
which is not sound.

Where an object with some fields is required, an object having these fields and also some others is allowed, according to the Liskov substitution principle \cite{DBLP:journals/toplas/LiskovW94}. One way of achieving this would be allowing weakening by forgetting fields:
\[X\matchess{m:u,\somefields} \vdc X\matchess{\somefields}\]
Alas, this is not sound either, since it allows to infer
$X\matchess{};\Gamma\vd x.m = 1 : \kint ;X\matchess{}$
and, by monotonicity, 
\[ X\matchess{m:\kstr};\Gamma\vd x.m = 1 : \kint ;X\matchess{m:\kstr}. \]
This hints to the fact that our $\vdc$ (defined over relation $\matches$) does not coincide with subtyping. Subtyping (at least in width) is nevertheless essential in an object-oriented setting and we actually permit it in function calls (see the explanation about rule 
$\fapprule$ in Section~\ref{sec:typingrules} and examples in Section~\ref{sec:examples}).

Equalising branches' postconditions might be done using
union types: if an attribute has type $t_1$ after one branch, and
$t_2$ after the other, we say it has type $t_1\lor t_2$. However, we
need to take special care; when trying to handle the case where an
attribute is set in one branch of the conditional, e.g.,
\begin{lstlisting}
  if (b) then x.m = 1 else 0
\end{lstlisting}
it may be tempting to use a weakening schema similar to
\[ \{\overline{n:t}\} \vdc \{\overline{n:t},m:u\lor\bot\}
   \quad m\not\in\overline{n}
\]

This turns out to be unsound, too, as shown by the following:
\begin{lstlisting}
  func(x) {ifhasattr(x, m) then x.m + 1 else 0}
\end{lstlisting}
 Using the weakening schema above, we can give it the type
 \[ [X;X\matchess{}] \To [ \kint; X \matchess{}]\]
 whereas calling this function with an argument containing field $m:\kstr$ 
 leads to a crash. 

\fi 

Therefore we propose a notion of type weakening as formulated in
Figure~\ref{fig:constraints-order}. This allows us to avoid the pitfall presented previously and 
give the function mentioned there a correct type
 \[ [X;X\matchess{m:\bot}] \To [ \kint; X \matchess{m:\bot\lor\kint}]\]
which ensures that the field $m$ is absent from its argument.


\vdcfigure

\subsection{Typing Rules}
\label{sec:typingrules}

The typing rules of our system are presented in Figure~\ref{fig:rules}.
A freshly created object has no fields, hence the form of rule 
$\newrule$. We impose an injective map from the set of type variables present
in the program to memory locations, therefore the type variable needs to be fresh.
The consequence is that
any relevant type variable occurring in the postcondition, but not in
the precondition of a judgement, refers to an object created within the expression
under consideration.
More precisely, whenever
$$
   \Psi_1; \Gamma \vd e : t; \Psi_2,
   \qquad X\in\dom{\Psi_2}\setminus\dom{\Psi_1, \Gamma} 
$$
$X$ is the type of an object created within $e$ (or phantom). Then we also know that all its fields not mentioned in the postcondition for $X$ are definitely absent, which is why the rule $\botrule$ is sound.

Rule $\raccrule$ governs field access. A field is accessible from an object (value) if the field's type is a type belonging to the set $\Types$.
Intuitively, a field can be accessed only if its type does not contain type $\bot$, that is, the field is actually present in the object.

Rule $\urule$ describes field update and works whether the field $m$ is
already present in the object or not. The postcondition is updated
accordingly, by using the operation $\updconstr$ from
Figure~\ref{fig:update-constraints}. The constraint related to $X$ in the
postcondition will record either the presence of a new field, or the (possible)
change of type of an already present fields (notice that most of the rules
defining $\updconstr$ are for the propagation of additions/changes and for
bookkeeping).
\updatefigure

The $\klet$ instruction provides a form of sequencing and the rule $\letrule$ types it accordingly.
%
We use the following notation:
\begin{displaymath}
  \begin{array}{rl}
    \klet\;x = e;\;es &\mapsto \letin{x}{e}{es}\\
    e;\; es           &\mapsto \letin{\_}{e}{es}
  
  \end{array}
\end{displaymath}

\iffalse
Rule $\condrule$ types our conditional expression. Following Hoare, the rule requires that both branches have the same postconditions.

Particular care is given to type the introspection operation $\kha$.
Rule $\ifhatrule$, following Hoare again, requires both branches to have the same
postconditions. It types an introspection expression against two different sets
of constraints, one \emph{assuming} the presence of the attribute $a$, the other
one \emph{assuming} the absence of the attribute $a$ in the object $v$. If one
of the two corresponding premises fail, the introspection expression may be
typed against rule $\ifhatrulethen$ or rule $\ifhatruleelse$, \emph{requiring}
that the attribute $a$ is present, respectively not present, in the object.
Intuitively, the first rule will typecheck when there is no static knowledge
about the presence/absence of the field $a$, while the other two will typecheck
when it is possible to determine it statically. Note that the necessary
information to discriminate is in the preconditions. These rules make use of a
definiteness update operation on constraints $\Psi[X\gets^+ \{a\}]$ (to assume or
to require the presence of $a$ in rule $\ifhatrule$ and rule $\ifhatrulethen$,
respectively) and $\Psi[X\gets^- \{a\}]$ (to assume the absence of $a$ in rule
$\ifhatruleelse$), presented in Figure~\ref{fig:definitness-update}. Note that $t^+$ is a partial operation and $\bot^+$ is undefined.
\fi

\typingrulesfiguresize{\normalsize}

Rule $\fdeclrule$ types a function declaration. It checks the body against the
given preconditions and postconditions, moreover the type is generalised on all
possible type variables not appearing in the typing context $\Gamma$ (in an
ML-style). This is done to abstract from the choices of type variable names
in types of the objects passed as arguments, as well as objects
created in the function body, while protecting the type variables referred to by
nonlocal identifiers.

The process of matching formal and actual parameters and preconditions can be
seen in the rule $\fapprule$. This rule, given a well-typed function
declaration: (\emph{i}) checks the actual parameters against the formal
parameters' types; (\emph{ii}) checks that the state at the call site (described
by $\Psi$) ensures the callee precondition ($\Psi_s$).
Note that this is expressed in terms of updates, because  each declared
function can be seen as a state updater, as well as inclusion of
domains, because the precondition cannot introduce new type variables (that
directly correspond to locations on heap).
%
All checks here are done modulo a renaming $\theta$ of type variables establishing a
one-to-one correspondence between formal and actual arguments and preconditions,
which is formally expressed with the $\parallel$ operator.
The other side condition ensures the type variables chosen for the objects created by the function are fresh.
%
Finally, we stipulate that the result type is the formal result type with type
variables renamed according to $\theta$, that adapts the type of the function
to the site of the function call. The final state corresponds here to
the initial state updated according to the callee postcondition $\Psi_t$.

Renaming is also an instrument we use to deal with aliasing and works together
with intersection types and their related rules, $\fcdecomp$ and
$\fdcomp{\mbox{i}}$. See the example in
Section~\ref{sec:example-intersection}
for an account on how renaming and intersection types work to deal with aliasing scenarios.

\defupdatefigure

Rule $\ifhatrule$ types an introspection expression against two different sets
of constraints, \emph{assuming} the presence (resp. absence) of the attribute $n$.
The rule $\ifhatrulete$ is a special case applicable if presence of
$n$ can be determined statically.


\iffalse
Rules $\vdcrule$ and $\typevdcrule$ are the means to apply weakening (explained
in detail in Section~\ref{sec:weakening}), to sets of constraints and to types,
respectively. It is worth noticing that it is by weakening that we perform
union-introduction (on fields' types), while union-elimination is done via rules
$\ifhatrule$ (a limited form, as we eliminate a disjunct only if it is $\bot$),
as well as via rules for base operations such as $\plusrule$.
A form of $\bot$-introduction is in rules $\botrule$ and $\ifhatruleelse$, while
there is no explicit elimination-rule for $\bot$, however this is implicit in
the rule $\urule$, as a previously absent field can be added.
\fi

%

\section{Expressivity of the System}
\label{sec:examples}

Let us now look at some examples that illustrate the strength of the
type system we propose.





An interesting point is what happens when an object is
modified and/or created inside a conditional instruction. We present four examples:
one where the same attribute is set in both branches, 
one where the assignment happens in one branch only, one in which an object is created and assigned to a field in one branch only, and one where object creation happens in both branches.
We assume we have a variable
called \texttt{hasarg} of type \texttt{bool} and a variable called \texttt{arg}
of type \texttt{string}.

\myparagraph{Setting the same attribute in both branches.}
Consider the code.
\begingroup
\lstset{texcl=true}
\begin{lstlisting}
     let x = new in
     // $x:X; X\matchess{}$
     if (ha) then x.m = a else x.m = "help"
     // $x:X; X \matchess{m:\kstring}$
\end{lstlisting}
\endgroup
We can type this example as follows:
\[ \infer{
   \Psi_1;\Gamma_1 \vd
      \ite{ha}
          {\mathit{x.m = a}}
          {\mathit{x.m = \mathtt{"help"}}}
         : \kstring; \Psi_2} 
{
  \begin{array}{c}
 \Psi_1;\Gamma_1\vd ha : \kbool
\\ 
 \Psi_1;\Gamma_1\vd x.m = a : \kstring;\Psi_2
\\
 \Psi_1;\Gamma_1\vd x.m = \mathtt{"help"} : \kstring;\Psi_2
\end{array}
}
\]
where
\begin{align*}
\Gamma_0 &= \{ ha : bool,a:\kstring \} \\
\Gamma_1 &= \Gamma_0,x:X \\
\Psi_1 &= X\matchess{} \\
\Psi_2 &= X\matchess{m:\kstring}
\end{align*}
Note that $\Psi_2 = \Psi_1 \updconstr X\matchess{m:\kstring}$ and, by using
the $\urule$ rule, we can derive
\[
\infer{
\Psi_1;\Gamma_1\vd x.m = s : \kstring;\Psi_2
}
{
\Psi_1;\Gamma_1\vd x : X
&
\Psi_1;\Gamma_1\vd s: \kstring
}
\]
for $s\equiv a$ as well as $s\equiv\mathtt{"help"}$.

\myparagraph{Setting an attribute in one branch only.}
Consider the code.
\begingroup
\lstset{texcl=true}
\begin{lstlisting}
     let x = new in
     // $x:X; X \matchess{m:\bot}$
     if (ha) then x.m = a else ""
     // $x:X; X\matchess{m:\kstring \lor\bot}$
\end{lstlisting}
\endgroup
\removelastskip\smallskip
\noindent
Typing the \texttt{then} branch looks like:
\[
\infer[(1)]{\Psi_1;\Gamma_1\vd\kdo{x.m=a}:\kstring;\;\Psi_2}
{
\infer{
 \begin{array}{c}
\Psi_1;\Gamma_1\vd  \kdo{x.m=a}:\kstring;\;X\matchess{m:\kstring}
 \\ \{m:\kstring\}\vdc\{m:\kstring\lor\bot\}
 \end{array}
 }
 {\Psi_1;\Gamma_1\vd x.m=a:\kstring;\;X_m\matchess{m:\kstring}}
}
\]
Typing the \texttt{else} branch looks like:
\[
\infer[(2)]
  { \Psi_1;\;\vd \mathtt{""} : \kstring;\;X\matchess{\mathit{m}:\kstring\lor\bot}}
  {
    \infer{\{m:\bot\}\vdc\{m:\kstring\lor\bot\}
          }
          {\kstring\vdc\kstring\lor\bot&\{\}\vdc\{\}}
  }
\]
Then, by putting the two branches together, we get:
\[
\infer{\Psi_1;\Gamma_1\vd\ite{ha}{\kdo{x.m=a}}{\mathtt{""}}:\kstring;\;\Psi_2}
{ 
 \Psi_1;\Gamma_1\vd ha : \kbool;\Psi_1
& 
 (1) & (2) 
}
\]
where
\begin{align*}
\Gamma_0 &= \{ ha : bool,a:\kstring \} \\
\Gamma_1 &= \Gamma_0,x:X \\
\Psi_1 &= X\matchess{m:\bot} \\
\Psi_2 &= X\matchess{m:\kstring\lor\bot}
\end{align*}


To type the whole $\klet$ we need to derive
\[
\infer{\vd \knew : X;X\matchess{m:\bot}}
{\vd\knew  : X;X\matchess{}}
\]
which is easily done using the (new) and (bot) rules; 
the side-conditions of (bot) are obviously respected here, however
they are needed to prevent the problems with weakening described in Section~\ref{sec:weakening}.
\par\bigskip\goodbreak

\myparagraph{Creating an object  in one branch only.}
Let \texttt{b} be of type \texttt{bool}, and \texttt{x} an object not containing field \texttt{a}. Consider:
\begin{lstlisting}
     if b then  
       x.a = new; 0
     else 0
\end{lstlisting}
\noindent %
Typing the \texttt{then} branch (with type Y for the new object) looks like:

\[
\infer{\Psi;\Gamma\vd  \kdo{x.a = \knew;0} : \kint; X\matchess{a:Y\lor\bot},Y\matchess{}}
{\Psi;\Gamma\vd  \kdo{x.a = \knew;0} : \kint; X\matchess{a:Y},Y\matchess{}}
\]

by applying rule $\vdcrule$, with $ \Gamma = \{b:\kbool,x:X\}$, $\Psi = \{X\matchess{a:\bot}\}$.

Similarly, for the \texttt{else} branch we want to prove
\[ \Psi;\Gamma\vd 0 : \kint; X\matchess{a:Y\lor\bot},Y\matchess{} \]

\noindent
We can easily infer
$\Psi;\Gamma\vd 0 : \kint; \Psi$.
Using rule $\vdccevolve$ we get
\[ X\matchess{a:\bot} \vdc X\matchess{a:\bot},Y\matchess{}\]
Then with some applications of $\vdc$ bookkeeping rules we can get
\[\Psi \vdc X\matchess{a:Y\lor\bot},Y\matchess{} \]
which leads us to the desired conclusion.

\begin{wrapfigure}[9]{r}[0em]{110pt}

\vspace{-2ex}

\begin{minipage}{100pt}
Input:\\
\includegraphics{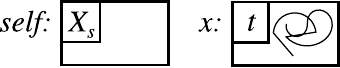}
\end{minipage}\\[1ex]

\begin{minipage}{100pt}
Output:\\
\phantom{self}$\;$\includegraphics{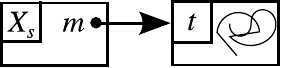}
\end{minipage}\\[1ex]

{\refstepcounter{figure}{\bf Figure \thefigure:} Graph fragments for input and output of the
  function type in the example.} 
\label{fig:fun-io}
\end{wrapfigure}
\removelastskip
\myparagraph{Function declaration and application.}
\label{sec:ex-fun-simple}
%
%
Consider a function that adds a field named $m$ with a value provided as
its second argument (of an arbitrary type $t$) to an object being its first argument:
\begin{lstlisting}
     func(self,x) { self.m = x }
\end{lstlisting}
\noindent
\def\tadd{t_{\mathit{add}}}
Let
$ \tadd = \forall X_s.\atype{X_s\matches\{\}}{X_s,t}{t}{X_s\matches\{m:t\}}.$
Observe now that the type ensures an important property of the object
graph 
in the heap that holds each time the function is called, therefore it is
invariant. 
 The type $\forall \vec{X}.[\ldots;\Psi_1]\To[\ldots;\Psi_2]$ may be
 read as ``for all graphs such that $\Psi_1$ holds before the call, $\Psi_2$ holds afterwards''.
In Figure~7, 
the boxes represent objects in the heap and have names
($X_s, t$) that stem from the types. In addition they are marked
with variables that reference them ($\mathit{self}, x$). 
The scribbles in the boxes hide the types that such an object can assume during the computation. The initial
input graph does not have an explicit connection between $X_s$ and $t$, but
in the result there is such a connection, from the now explicit field $m$ to $t$. This is a simple
example, but the invariants in real programs may involve complicated
graphs that can be expressed straightforwardly in this way.

We can derive $t_{\mathit{add}}$,  by rule $\fdeclrule$
\[
\infer{\vd\fdecl{\mathit{self},x}{ t_{\mathit{add}}}
             {\ldots} : t_{\mathit{add}};\emptyset}
{
    {\mathit{self}:X_s,x:t;X_s\matches\{\}\vd \mathit{self}.m=x:t;X_s\matches\{m:t\}}
}
\]

Now we apply the above function to a newly created object:
\begin{lstlisting}
     let init = func(self,x) { self.m = x }
     let o = new // o : Xo; Xo <# {}
     init(o,42)  // o : Xo; Xo <# {m:int}
\end{lstlisting}
\noindent
The renaming $\theta$ connects the formal and actual parameters,
sending $X_s$ to $X_o$;
putting
\begin{align*}
  \Gamma_0&=\{o:X_o;\mathit{init} : \tadd\}, \qquad
   \Psi_0 = X_o\matchess{} 
\end{align*}
We can infer, by $\fapprule$,
\[ \Psi_0;\;\Gamma_0\vd 
        \mathit{init}(o,42):\kint;\;\Psi_0\updconstr\theta(X\matchess{m:\kint})
\]

Here, we can see our form of subtyping in width at work (see Section~\ref{sec:weakening} and rule $\fapprule$);
observe that the same function may be called on an object containing some fields
already: if $\Psi_0 = X_o\matchess{n:u}$, then
$\Psi_0 \updconstr \theta(X_s\matchess{}) = \Psi_0$.
In both cases, the renaming is a witness that the program state satisfies the function precondition.

\myparagraph{Intersection types.}
\label{sec:example-intersection}
As hinted, the idea behind intersection types in our system is that they capture
allowed aliasing scenarios. Consider the following function:

\begin{lstlisting}
     func(x,y) { x.m = 1; y.m }
\end{lstlisting}
This function can work in either of the following scenarios:
\begin{itemize}\topsep 1pt plus 2pt minus 1pt%
\item[(i)] the actual parameter for $y$ has field $m$ before it is passed to the function,
\item[(ii)] the actual parameters for $x$ and $y$ are the same object.
\end{itemize}

Hence the type of the function will be written as $t_1\land t_2$,
where $t_i$ represent types corresponding to  scenarios (i) and (ii):
\[ t_1 = \forall X,Y.[X,Y;\Psi_1]\implies [u;\Psi_1\updconstr X\matchess{m:\kint}]\]
with
$\Psi_1 = X\matchess{},Y\matchess{m:u}, $ and
\[ t_2 = \forall
X.[X,X;X\matchess{}]\implies[\kint;X\matchess{m:\kint}]\]

In practice one does not need to write intersection types, but instead write multiple contracts for a function (and add more as needed), for example (with a fair dose of syntactic sugar):
\begin{lstlisting}
     f : [X,Y;Y.m:U] => [U;X.m:int]
     f : [X,X] => [int;X.m:int]
     func f(x,y) { x.m = 1; y.m }
\end{lstlisting}

\ifcolorpoint
\myparagraph{Point, ColorPoint.}
The following example is an encoding of a paradigmatic example in our system. The type of the $mv$ method is a function type that takes as a parameter an object containing at least a field $x$ of type \texttt{int}. Note that this is an imperative version (without MyType) of an analogous example in \cite{FisherHM94}.

\begin{lstlisting}
let o = new;
o.x = 7;    // r = {x:int}
// Tmv = forall Xs.[Xs,int;Xs<#r] => [Xs,Xs<#r]
o.mv = func(self,dx){ self.x = self.x+dx; self }
// o : Xo; Xo <# { x:int, mv:Tmv } 
o.c = "blue";
// o : Xo; Xo <# { x:int, mv:Tmv, c:string } 
o.mv(o,3); // can call mv: Xo <# { x:int } holds
o.c        // we can still read the field "c"
\end{lstlisting}
\noindent
In this example, we again see our subtyping in width at work.
The method $mv$ requires an object with a field $x$, but it works also if the actual parameter contains extra fields, in our case field $c$. Moreover, the field $c$ is still accessible after the method call.
\fi 

\section{Conclusions and future work}
\label{sec:conclusions}



Our contribution is a novel type system
for typing dynamic languages in a
retrofitted manner, 
with particular emphasis over the flow of control and with the
aim of tracing the changes of object interfaces at runtime.
We express this change by means of Hoare-like triples that describe the
structure of relevant objects before an expression is executed and after its
execution.
The type reconstruction for our system
seems undecidable (in fact, a slightly weakened version of the
intersection type system can be embedded in the language \cite{KfouryW2004}).
However, there is a strong evidence that the type system becomes decidable
when type annotations are provided for functions (by a programmer or by
a non-complete heuristics), as seen for similar annotation
schemes \cite{JayJ08}. 

As a further development, we would like to apply \Lucretia's approach
to regulate control flow in JavaScript. Moreover, we want to couple our system
with the \emph{gradual typing} method
\cite{DBLP:conf/oopsla/BrachaG93,DBLP:journals/entcs/AndersonD03,DBLP:conf/ecoop/SiekT07},
that sup\-ports evolving an untyped program into a typed one, possibly by using the
like-type approach of~\cite{WrigstadNLOV10b}.

\bibliographystyle{eptcs}
\bibliography{refl-short-doi}

\begin{thebibliography}{10}
\providecommand{\bibitemdeclare}[2]{}
\providecommand{\surnamestart}{}
\providecommand{\surnameend}{}
\providecommand{\urlprefix}{Available at }
\providecommand{\url}[1]{\texttt{#1}}
\providecommand{\href}[2]{\texttt{#2}}
\providecommand{\urlalt}[2]{\href{#1}{#2}}
\providecommand{\doi}[1]{doi:\urlalt{http://dx.doi.org/#1}{#1}}
\providecommand{\bibinfo}[2]{#2}

\bibitemdeclare{book}{DBLP:books/daglib/0098135}
\bibitem{DBLP:books/daglib/0098135}
\bibinfo{author}{Torben \surnamestart Amtoft\surnameend},
  \bibinfo{author}{Hanne~Riis \surnamestart Nielson\surnameend} \&
  \bibinfo{author}{Flemming \surnamestart Nielson\surnameend}
  (\bibinfo{year}{1999}): \emph{\bibinfo{title}{Type and effect systems -
  behaviours for concurrency}}.
\newblock \bibinfo{publisher}{Imperial College Press}, \doi{10.1142/p132}.

\bibitemdeclare{article}{DBLP:journals/entcs/AndersonD03}
\bibitem{DBLP:journals/entcs/AndersonD03}
\bibinfo{author}{Christopher \surnamestart Anderson\surnameend} \&
  \bibinfo{author}{Sophia \surnamestart Drossopoulou\surnameend}
  (\bibinfo{year}{2003}): \emph{\bibinfo{title}{{BabyJ}: from object based to
  class based programming via types}}.
\newblock {\sl \bibinfo{journal}{ENTCS}}
  \bibinfo{volume}{82}(\bibinfo{number}{8}), pp. \bibinfo{pages}{53--81},
  \doi{10.1016/S1571-0661(04)80802-8}.

\bibitemdeclare{article}{BarendregtCD83}
\bibitem{BarendregtCD83}
\bibinfo{author}{Henk \surnamestart Barendregt\surnameend},
  \bibinfo{author}{Mario \surnamestart Coppo\surnameend} \&
  \bibinfo{author}{Mariangiola \surnamestart Dezani-Ciancaglini\surnameend}
  (\bibinfo{year}{1983}): \emph{\bibinfo{title}{A Filter Lambda Model and the
  Completeness of Type Assignment}}.
\newblock {\sl \bibinfo{journal}{J. Symbolic Logic}}
  \bibinfo{volume}{48}(\bibinfo{number}{4}), pp. \bibinfo{pages}{931--940},
  \doi{10.2307/2273659}.

\bibitemdeclare{inproceedings}{DBLP:conf/oopsla/BrachaG93}
\bibitem{DBLP:conf/oopsla/BrachaG93}
\bibinfo{author}{Gilad \surnamestart Bracha\surnameend} \&
  \bibinfo{author}{David \surnamestart Griswold\surnameend}
  (\bibinfo{year}{1993}): \emph{\bibinfo{title}{Strongtalk: Typechecking
  {S}malltalk in a Production Environment}}.
\newblock In: {\sl \bibinfo{booktitle}{Proc. of OOPSLA'93}}, pp.
  \bibinfo{pages}{215--230}, \doi{10.1145/165854.165893}.

\bibitemdeclare{article}{FisherHM94}
\bibitem{FisherHM94}
\bibinfo{author}{Kathleen \surnamestart Fisher\surnameend},
  \bibinfo{author}{Furio \surnamestart Honsell\surnameend} \&
  \bibinfo{author}{John~C. \surnamestart Mitchell\surnameend}
  (\bibinfo{year}{1994}): \emph{\bibinfo{title}{A lambda calculus of objects
  and method specialization}}.
\newblock {\sl \bibinfo{journal}{Nord. J. Comp.}} \bibinfo{volume}{1}, pp.
  \bibinfo{pages}{3--37}.

\bibitemdeclare{inproceedings}{Gifford86}
\bibitem{Gifford86}
\bibinfo{author}{David~K. \surnamestart Gifford\surnameend} \&
  \bibinfo{author}{John~M. \surnamestart Lucassen\surnameend}
  (\bibinfo{year}{1986}): \emph{\bibinfo{title}{Integrating functional and
  imperative programming}}.
\newblock In: {\sl \bibinfo{booktitle}{Proc. of LFP'89}}, \bibinfo{series}{LFP
  '86}, \bibinfo{publisher}{ACM}, \bibinfo{address}{New York, NY, USA}, pp.
  \bibinfo{pages}{28--38}, \doi{10.1145/319838.319848}.

\bibitemdeclare{inproceedings}{JayJ08}
\bibitem{JayJ08}
\bibinfo{author}{Barry \surnamestart Jay\surnameend} \&
  \bibinfo{author}{Simon~Peyton \surnamestart Jones\surnameend}
  (\bibinfo{year}{2008}): \emph{\bibinfo{title}{Scrap Your Type Applications}}.
\newblock In: {\sl \bibinfo{booktitle}{Proceedings of the 9th international
  conference on Mathematics of Program Construction}}, \bibinfo{series}{LNCS},
  \bibinfo{publisher}{Springer-Verlag}, \bibinfo{address}{Berlin, Heidelberg},
  pp. \bibinfo{pages}{2--27}, \doi{10.1007/978-3-540-70594-9\_2}.

\bibitemdeclare{article}{KfouryW2004}
\bibitem{KfouryW2004}
\bibinfo{author}{A.J. \surnamestart Kfoury\surnameend} \& \bibinfo{author}{J.B.
  \surnamestart Wells\surnameend} (\bibinfo{year}{2004}):
  \emph{\bibinfo{title}{Principality and type inference for intersection types
  using expansion variables}}.
\newblock {\sl \bibinfo{journal}{Theor. Comput. Sci.}}
  \bibinfo{volume}{311}(\bibinfo{number}{1--3}), pp. \bibinfo{pages}{1--70},
  \doi{10.1016/j.tcs.2003.10.032}.

\bibitemdeclare{article}{KoMCA06}
\bibitem{KoMCA06}
\bibinfo{author}{Andrew~J. \surnamestart Ko\surnameend},
  \bibinfo{author}{Brad~A. \surnamestart Myers\surnameend},
  \bibinfo{author}{Michael~J. \surnamestart Coblenz\surnameend} \&
  \bibinfo{author}{Htet~Htet \surnamestart Aung\surnameend}
  (\bibinfo{year}{2006}): \emph{\bibinfo{title}{An Exploratory Study of How
  Developers Seek, Relate, and Collect Relevant Information during Software
  Maintenance Tasks}}.
\newblock {\sl \bibinfo{journal}{IEEE Transactions on Software Engineering}}
  \bibinfo{volume}{32}(\bibinfo{number}{12}), pp. \bibinfo{pages}{971--987},
  \doi{10.1109/TSE.2006.116}.

\bibitemdeclare{article}{DBLP:journals/toplas/LiskovW94}
\bibitem{DBLP:journals/toplas/LiskovW94}
\bibinfo{author}{Barbara \surnamestart Liskov\surnameend} \&
  \bibinfo{author}{Jeannette~M. \surnamestart Wing\surnameend}
  (\bibinfo{year}{1994}): \emph{\bibinfo{title}{A Behavioral Notion of
  Subtyping}}.
\newblock {\sl \bibinfo{journal}{ACM Trans. Program. Lang. Syst.}}
  \bibinfo{volume}{16}(\bibinfo{number}{6}), pp. \bibinfo{pages}{1811--1841},
  \doi{10.1145/197320.197383}.

\bibitemdeclare{inproceedings}{Marino09}
\bibitem{Marino09}
\bibinfo{author}{Daniel \surnamestart Marino\surnameend} \&
  \bibinfo{author}{Todd \surnamestart Millstein\surnameend}
  (\bibinfo{year}{2009}): \emph{\bibinfo{title}{A generic type-and-effect
  system}}.
\newblock In: {\sl \bibinfo{booktitle}{Proceedings of the 4th International
  Workshop on Types in Language Design and Implementation}},
  \bibinfo{series}{TLDI '09}, \bibinfo{publisher}{ACM}, \bibinfo{address}{New
  York, NY, USA}, pp. \bibinfo{pages}{39--50}, \doi{10.1145/1481861.1481868}.

\bibitemdeclare{inproceedings}{MayerHRTS12}
\bibitem{MayerHRTS12}
\bibinfo{author}{Clemens \surnamestart Mayer\surnameend},
  \bibinfo{author}{Stefan \surnamestart Hanenberg\surnameend},
  \bibinfo{author}{Romain \surnamestart Robbes\surnameend},
  \bibinfo{author}{\surnamestart \'{E}ric Tanter\surnameend} \&
  \bibinfo{author}{Andreas \surnamestart Stefik\surnameend}
  (\bibinfo{year}{2012}): \emph{\bibinfo{title}{An empirical study of the
  influence of static type systems on the usability of undocumented software}}.
\newblock In: {\sl \bibinfo{booktitle}{Proceedings of OOPSLA}},
  \bibinfo{publisher}{ACM}, pp. \bibinfo{pages}{683--702},
  \doi{10.1145/2384616.2384666}.

\bibitemdeclare{article}{Prechelt00}
\bibitem{Prechelt00}
\bibinfo{author}{Lutz \surnamestart Prechelt\surnameend}
  (\bibinfo{year}{2000}): \emph{\bibinfo{title}{An Empirical Comparison of
  Seven Programming Languages}}.
\newblock {\sl \bibinfo{journal}{Computer}}
  \bibinfo{volume}{33}(\bibinfo{number}{10}), pp. \bibinfo{pages}{23--29},
  \doi{10.1109/2.876288}.

\bibitemdeclare{article}{Sasso96}
\bibitem{Sasso96}
\bibinfo{author}{William \surnamestart Sasso\surnameend}
  (\bibinfo{year}{1996}): \emph{\bibinfo{title}{Cognitive processes in program
  comprehension: An empirical analysis in the Context of software
  reengineering}}.
\newblock {\sl \bibinfo{journal}{Journal of Systems and Software}}
  \bibinfo{volume}{34}(\bibinfo{number}{3}), pp. \bibinfo{pages}{177--189},
  \doi{10.1016/0164-1212(95)00074-7}.

\bibitemdeclare{inproceedings}{DBLP:conf/ecoop/SiekT07}
\bibitem{DBLP:conf/ecoop/SiekT07}
\bibinfo{author}{Jeremy~G. \surnamestart Siek\surnameend} \&
  \bibinfo{author}{Walid \surnamestart Taha\surnameend} (\bibinfo{year}{2007}):
  \emph{\bibinfo{title}{Gradual Typing for Objects}}.
\newblock In: {\sl \bibinfo{booktitle}{Proc. of ECOOP'07}}, pp.
  \bibinfo{pages}{2--27}, \doi{10.1007/978-3-540-73589-2\_2}.

\bibitemdeclare{misc}{Stephenson01}
\bibitem{Stephenson01}
\bibinfo{author}{Ed~\surnamestart Stephenson\surnameend}
  (\bibinfo{year}{2001}): \emph{\bibinfo{title}{Perl Runs Sweden's Pension
  System: A Fallback Application Built in Six Months Earns the Prime Role}}.
\newblock
  \bibinfo{howpublished}{\url{http://epq.com.co/softw_internet/trends/perl/eje%
mplo/}}.
\newblock \bibinfo{note}{(accessed October 31, 2014)}.

\bibitemdeclare{inproceedings}{WrigstadEFNV09}
\bibitem{WrigstadEFNV09}
\bibinfo{author}{Tobias \surnamestart Wrigstad\surnameend},
  \bibinfo{author}{Patrick \surnamestart Eugster\surnameend},
  \bibinfo{author}{John \surnamestart Field\surnameend}, \bibinfo{author}{Nate
  \surnamestart Nystrom\surnameend} \& \bibinfo{author}{Jan \surnamestart
  Vitek\surnameend} (\bibinfo{year}{2009}): \emph{\bibinfo{title}{Software
  hardening: a research agenda}}.
\newblock In: {\sl \bibinfo{booktitle}{Proceedings for the 1st Workshop on
  Script to Program Evolution}}, \bibinfo{series}{STOP '09},
  \bibinfo{publisher}{ACM}, \bibinfo{address}{New York, NY, USA}, pp.
  \bibinfo{pages}{58--70}, \doi{10.1145/1570506.1570513}.

\bibitemdeclare{inproceedings}{WrigstadNLOV10b}
\bibitem{WrigstadNLOV10b}
\bibinfo{author}{Tobias \surnamestart Wrigstad\surnameend},
  \bibinfo{author}{Francesco~Zappa \surnamestart Nardelli\surnameend},
  \bibinfo{author}{Sylvain \surnamestart Lebresne\surnameend},
  \bibinfo{author}{Johan \surnamestart \"{O}stlund\surnameend} \&
  \bibinfo{author}{Jan \surnamestart Vitek\surnameend} (\bibinfo{year}{2010}):
  \emph{\bibinfo{title}{Integrating typed and untyped code in a scripting
  language}}.
\newblock In: {\sl \bibinfo{booktitle}{Proc. of POPL'10}},
  \bibinfo{publisher}{ACM}, \bibinfo{address}{New York, NY, USA}, pp.
  \bibinfo{pages}{377--388}, \doi{10.1145/1706299.1706343}.

\end{thebibliography}

\end{document}
\endinput
\[
\infer{}
{}
\]